\documentclass[epjST]{svjour}
\usepackage{graphics}
\usepackage{color}
\usepackage{graphicx}
\begin{document}
\title{Updated sensitivity of DUNE in 3+1  scenario with far and near detectors}
\author{Monojit Ghosh\fnmsep\thanks{\email{monojit$\_$rfp@uohyd.ac.in}} \and Rukmani Mohanta\fnmsep\thanks{\email{: rmsp@uohyd.ac.in}} }
\institute{School of Physics, University of Hyderabad, Hyderabad - 500046, India}
\abstract{
In this paper we present the updated physics sensitivity of DUNE in presence of a light sterile neutrino with both far and near detectors. In the previous studies, the sensitivities were obtained using the configuration of DUNE as described in the conceptual design report (CDR). In this article, we consider the configuration of DUNE as given in the technical design report (TDR) and study the capability of this experiment to constrain the sterile mixing parameters as well as its capability to measure the standard oscillation parameters in 3+1 scenario. Our results show that in 3+1 scenario, the sensitivity of DUNE to measure the mass hierarchy, octant and CP violation deteriorates if we only consider the far detector. However,  a combined analysis of far and near detector improves the sensitivity. 
} 
\maketitle
\section{Introduction}
\label{intro}
In standard three flavour scenario, neutrino oscillation is described by three mixing angles: $\theta_{12}$, $\theta_{13}$ and $\theta_{23}$, two mass squared differences: $\Delta m^2_{21}$ and $\Delta m^2_{31}$ and one Dirac type CP violating phase: $\delta_{\rm CP}$. In the last two decades, neutrino oscillation experiments have measured some of the parameters with excellent precision \cite{Esteban:2020cvm}. At present the remaining unknowns in the three flavour scenario are: (i) neutrino mass hierarchy which can be either normal i.e., $\Delta m^2_{31} > 0$ or inverted i.e., $\Delta m^2_{31} < 0$, (ii) octant of the atmospheric mixing angle which can be either higher i.e., $\theta_{23} > 45^\circ$ or lower i.e., $\theta_{23} < 45^\circ$ and (iii) the CP violating phase $\delta_{\rm CP}$. The current trend of the data obtained from T2K \cite{T2K:2021xwb} and NO$\nu$A \cite{NOvA:2021nfi} experiments, show a mild preference towards normal hierarchy, higher octant and maximal CP violation i.e., $\delta_{\rm CP} = -90^\circ$. The aim of the future neutrino oscillation experiments is to establish these hints from the current generation experiments into firm footing. The proposed long-baseline experiment DUNE \cite{DUNE:2020ypp} in Fermilab is an example of such an experiment. 

Apart from the standard three flavour scenario, neutrino oscillation experiments provide an opportunity to study various new physics scenarios. One of such scenarios is the existence of light sterile neutrino in eV scale. Evidence from various experiments suggest that there can be a light sterile neutrino in eV scale. For a detailed review on that subject we refer to Ref. \cite{Abazajian:2012ys}. Sterile neutrinos are SU(2) singlets, which do not interact with the other Standard Model (SM) particles but they can undergo oscillation. If these kind of neutrinos exist in nature, then the standard neutrino oscillation probabilities are modified and therefore, one can observe signatures of such oscillations in the long-baseline experiments in both near and far detectors. As the oscillation of the sterile neutrinos are governed by a mass square difference around 1 eV$^2$, the oscillations will be averaged out at the far detector (FD), however, this can affect the measurement of the standard oscillation parameters. On the other hand, the near detector provides an opportunity to observe the oscillations of the sterile neutrinos directly.  In this article, we will study the effect of sterile oscillations in DUNE in presence of both ND and FD. Study of light sterile neutrinos in the context of DUNE has been studied in the past \cite{Berryman:2015nua,Gandhi:2015xza,Agarwalla:2016xxa,Agarwalla:2016xlg,Coloma:2017ptb,Choubey:2017cba,Choubey:2017ppj,Choubey:2016fpi,Gandhi:2017vzo}. But all these studies has been done using the older configuration of DUNE. In this article we update the sensitivity of DUNE in the presence of a light sterile neutrino i.e., 3+1 scenario, with the latest configuration of DUNE as described in the Technical Design Report (TDR) \cite{DUNE:2020ypp}. Note that in Ref. \cite{DUNE:2020fgq}, the DUNE collaboration has performed an analysis of 3+1 scenario with the updated configuration but they only presented the results of the bound on the sterile mixing parameters. On the other hand in this article, in addition to the bound on the sterile mixing parameters we will also demonstrate the effect of sterile oscillations on the measurement of standard oscillation parameters in presence of both FD and ND.

The paper is organized as follows. In the next section, we will briefly discuss the 3+1 scenario and the relevant new oscillation parameters. In Section \ref{sim} we will give the experimental and simulation detail which we consider in our analysis. In Section \ref{res} we present our results and finally in Section \ref{sum} we will summarize and conclude. 

\section{The 3+1 Scenario}
\label{sce}

In presence of a light sterile neutrino, the unitary PMNS matrix U can be parameterized by following:
\begin{eqnarray}
U =  U_{34}(\theta_{34},\delta_{34})
U_{24}(\theta_{24},\delta_{24})
U_{14}(\theta_{14},0)
U^{3\nu}\,, \\
U^{3\nu}
=
U_{23}(\theta_{23},0)
U_{13}(\theta_{13},\delta_{\rm CP})
U_{12}(\theta_{12},0)\,,
\end{eqnarray}
where $U_{ij}(\theta_{ij},\delta_{ij}\footnote{We will denote $\delta_{13}$ as $\delta_{\rm CP}$ throughout the paper.})$ denotes a rotation in the $(i,j)$-plane with mixing angle $\theta_{ij}$ and CP-violating phase $\delta_{ij}$ and $U^{3\nu}$ corresponds to the ordinary three-flavor leptonic mixing matrix. Therefore, we understand that in 3+1 scenario, we have three new mixing angles which are $\theta_{14}$, $\theta_{24}$ and $\theta_{34}$ and two new phases which are $\delta_{24}$ and $\delta_{34}$. Apart from these, we will also have a new mass square difference, we will call it as $\Delta m^2_{41}$.

\section{Experimental and Simulation Details}
\label{sim}

We simulate the experiment DUNE using the GLoBES software \cite{Huber:2004ka,Huber:2007ji}.  For DUNE, we use the official GLoBES files of the DUNE technical design report \cite{DUNE:2021cuw}.  We have considered a far detector of 40 kt liquid argon time-projection chamber detector is placed at a distance of 1300 km from the neutrino source having a power of 1.2 MW  delivering $1.1 \times 10^{21}$ POT per year with a running time of 7 years. The run-time is divided equally in neutrino mode and antineutrino mode. For the near detector, we have considered an identical detector as of the far detector but having a detector mass of 0.147 kt located at a distance of 0.574 km from the source \cite{DUNE:2021tad}. 

\section{Results}
\label{res}

To estimate the sensitivity, we calculate the statistical $\chi^2$ using
\begin{equation}
\chi^2_{{\rm stat}} = 2 \sum_{i=1}^n \bigg[ N^{{\rm test}}_i - N^{{\rm true}}_i - N^{{\rm true}}_i \log\bigg(\frac{N^{{\rm test}}_i}{N^{{\rm true}}_i}\bigg) \bigg]\,,
\end{equation}
where $n$ is the number of energy bins, $N^{{\rm true}}$ is the number of true events, and $N^{{\rm test}}$ is the number of test events. We incorporate the systematics by the method of pulls. In our analysis we have considered only charge current events and  as the charged current events are not sensitive to the parameter $\theta_{34}$, we have assumed the value of this parameter to be zero throughout our calculation. We have generated all the results assuming true hierarchy as normal. 
\begin{figure}
\begin{center}
\includegraphics[scale=0.4]{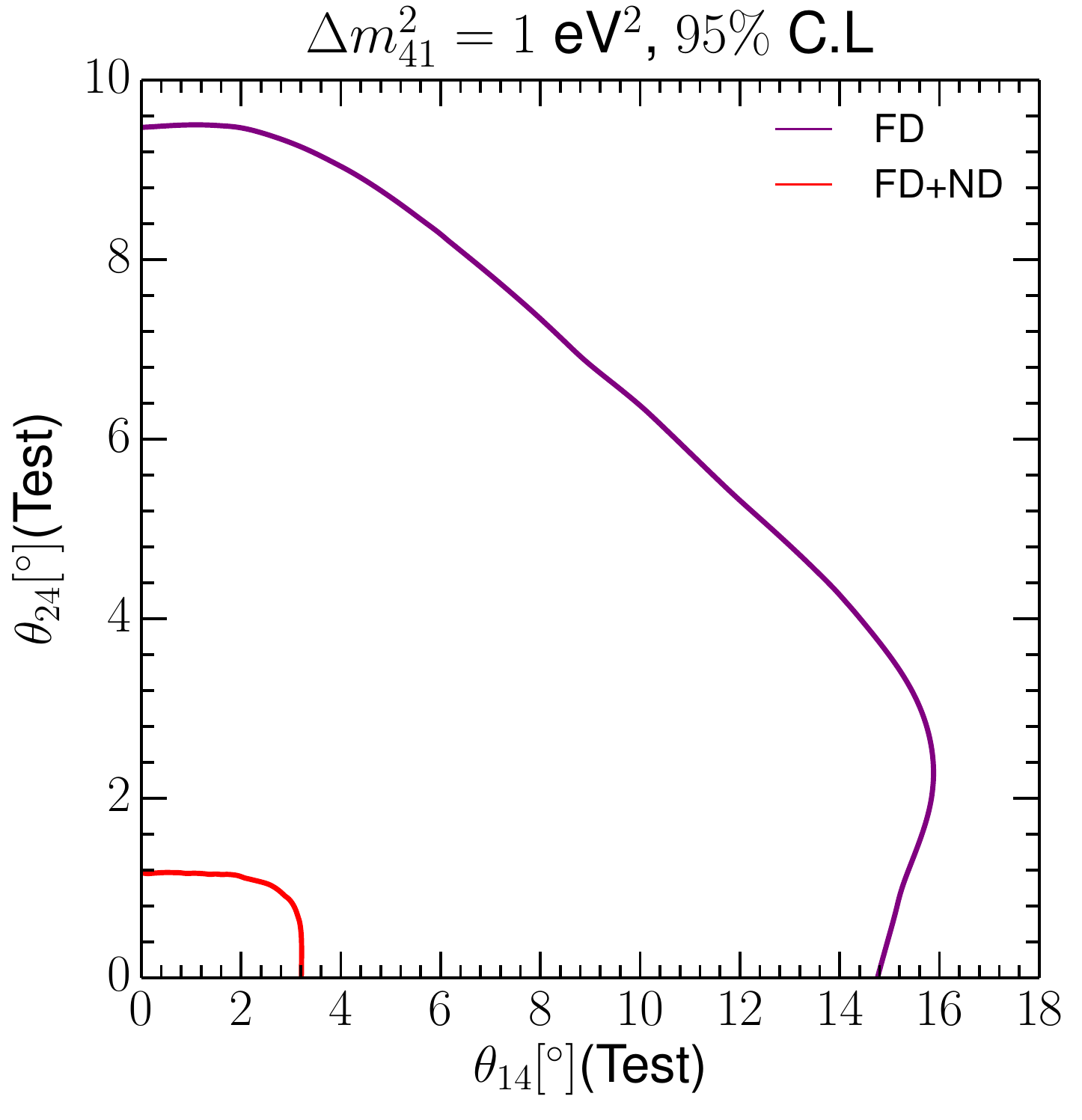}
\end{center}
\caption{$2\sigma$ bound on the sterile mixing angles assuming $\Delta m^2_{41} = 1$ eV$^2$.}
\label{fig:1}       
\end{figure}

In Fig. \ref{fig:1}, we have shown the $2 \sigma$ bound on the sterile mixing parameters $\theta_{14}$ and $\theta_{24}$ assuming the value of $\Delta m^2_{41}$ to be 1 eV$^2$. In generating this plot, we have considered the following true values of the parameters: $\sin^2\theta_{12} =  0.304$, $\sin^2\theta_{13} = 0.02219$, $\theta_{23} = 45^\circ$, $\delta_{\rm CP} = -90^\circ$, $\Delta m^2_{21} = 7.42 \times 10^{-5}$ eV$^2$ and $\Delta m^2_{31} = 2.517 \times 10^{-3}$ eV$^2$. The true value of all the sterile mixing parameters are zero. In the test we have minimzed the parameter $\theta_{23}$ between $40^\circ$ and $50^\circ$ and the parameters $\delta_{\rm CP}$ and $\delta_{24}$ between $-180^\circ$ and $180^\circ$. In this figure, the purple curve corresponds to the case when only FD is considered and the red curve corresponds to the case when both FD and ND are being considered. From this figure we understand that DUNE has an excellent capability to constrain the values of $\theta_{14}$ and $\theta_{24}$, if we include both FD and ND as compared to only FD. This is because, when we include ND, oscillations due to $\Delta m^2_{41}$ gets fully developed whereas at FD the oscillation due to $\Delta m^2_{41}$ gets averaged out. The result obtained in Fig. \ref{fig:1} is consistent with Ref. \cite{DUNE:2020fgq}.
\begin{figure}
\begin{center}
\includegraphics[scale=0.28]{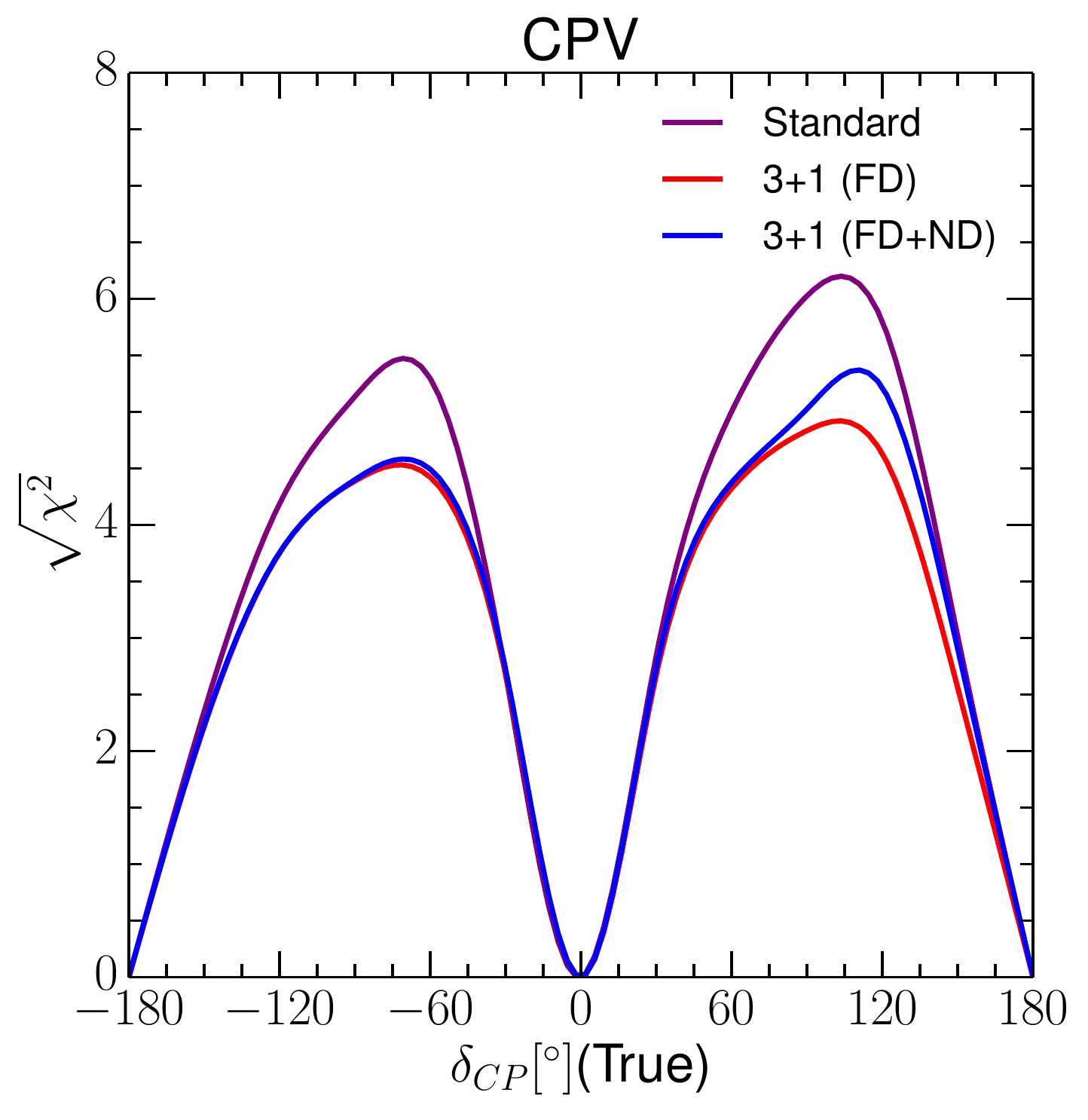}
\includegraphics[scale=0.28]{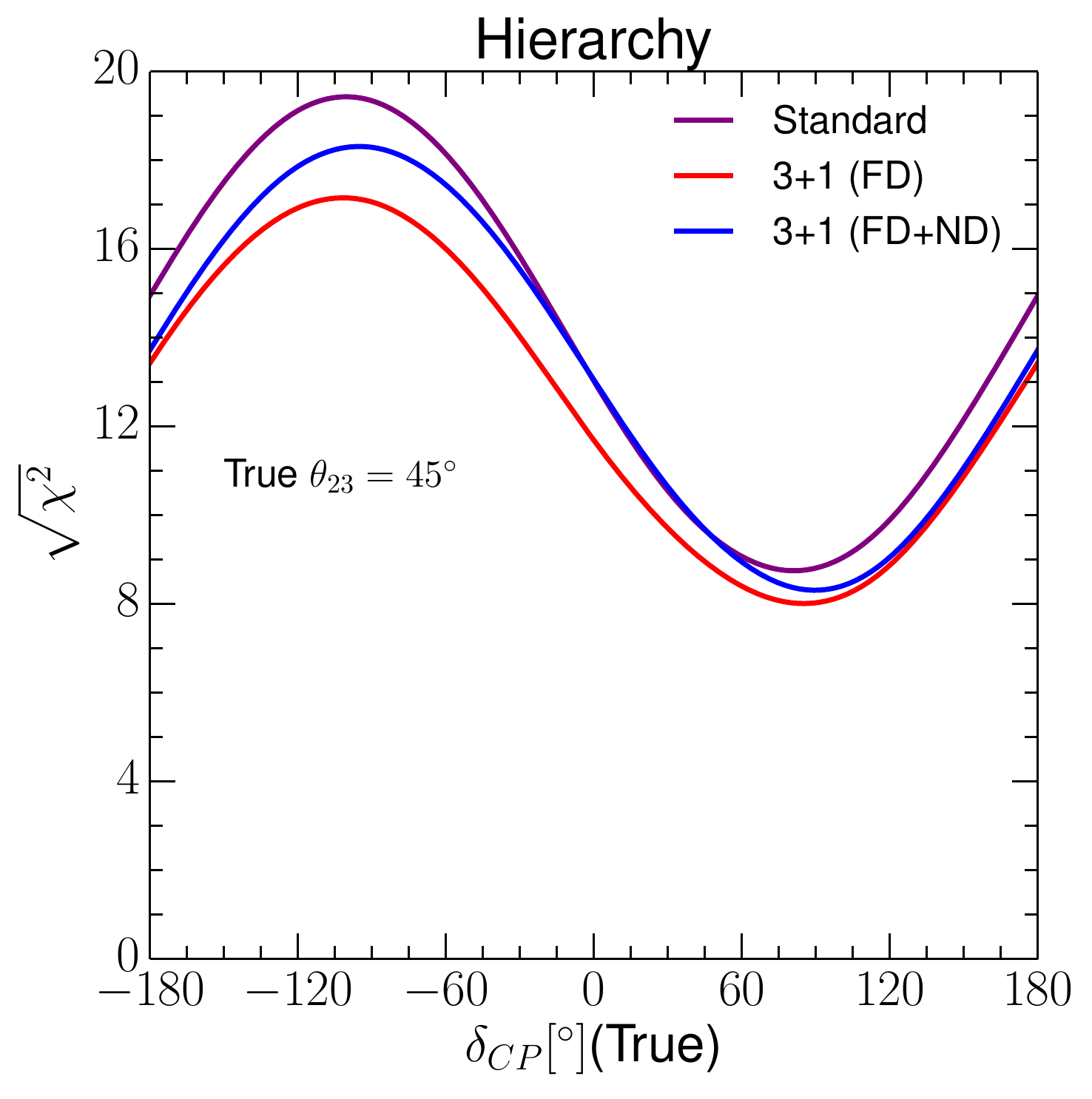}
\includegraphics[scale=0.28]{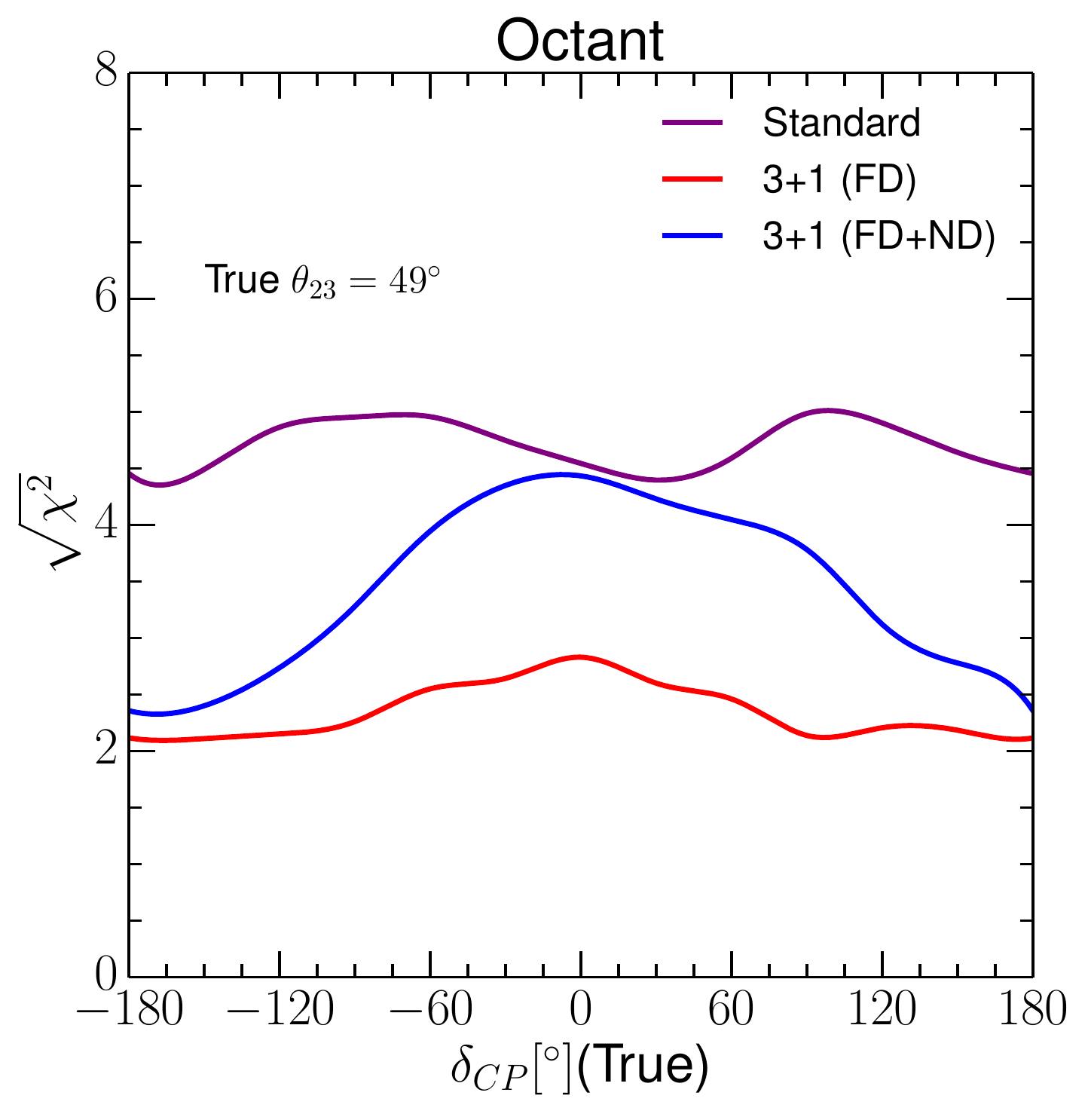} \\
\end{center}
\caption{Effect of ligh sterile neutrino in the measurement of standard oscillation parameters. }
\label{fig:2}       
\end{figure}

In Fig. \ref{fig:2}, we have presented the effect of light sterile neutrino in the measurement of standard oscillation parameters. The left, middle and right panels correspond to the CP violation discovery sensitivity, hierarchy sensitivity and octant sensitivity respectively. In each panel, the purple, red and blue curves correspond to the standard, sterile neutrino with only FD and sterile neutrino with FD+ND respectively. In these panels we have considered the true value of sterile parameters as $\theta_{14} = \theta_{24} = 5^\circ$ and $\delta_{24} = 0^\circ$. The mass square difference $\Delta m^2_{41}$ is kept fixed at 1 eV$^2$ in both true and test. We have minimized the parameters $\theta_{14}$, $\theta_{24}$ and $\delta_{24}$ in the test. For CP violation and hierarchy sensitivity we have considered true value of $\theta_{23}$ as $45^\circ$ whereas for the octant sensitivity we have considered the true value of $\theta_{23}$ as $49^\circ$. In all the three panel we notice that in presence of a light sterile neutrino, the sensitivity deteriorates as compared to the standard scenario. However, we notice that the sensitivity corresponding to FD+ND is better than the only FD in the 3+1 scenario. In fact, the hierarchy and octant sensitivity in 3+1 scenario for FD+ND, is almost same as that of the standard scenario for $\delta_{\rm CP} = 0^\circ$.  For CP violation sensitivity, FD+ND is slightly better than only FD in 3+1 scenario.

\section{Summary and Conclusion}
\label{sum}

In this article, we have updated the physics sensitivity of DUNE in the presence of a light sterile neutrino. The previous studies in this topic were performed using the configuration of DUNE as described in the CDR. In this paper, we have considered the configuration of DUNE as given in the TDR and updated the sensitivity of DUNE in measuring the standard oscillation parameters in presence of a light sterile neutrino including both FD and ND. Considering an identical near detector as of the far detector and taking the sterile mass squared difference to be 1 eV$^2$, we have shown that DUNE have excellent capability to constrain the sterile mixing angles if we combine the data from FD and ND. Our result is consistent with the results as given in Ref.  \cite{DUNE:2020fgq}. Further we have shown that in presence of a sterile neutrino, the CP violation, hierarchy and octant sensitivities deteriorate as compared to the standard scenario if we consider only FD. A combined analysis of FD and ND improves the sensitivity in the 3+1 scenario. For hierarchy and octant, the sensitivity in standard scenario is almost equal to the sensitivity in the 3+1 scenario for FD+ND if the true value of $\delta_{\rm CP}$ is around $0^\circ$. For CP violation, the sensitivity for FD+ND is slightly better than only FD in 3+1 scenario. The results obtained in this article show the importance of the near detector in measuring the standard oscillation parameters in presence of a light sterile neutrino.

\section*{Acknowledgements}
MG acknowledges Ramanujan Fellowship of SERB, Govt. of India,  through grant no: RJF/2020/000082. RM acknowledges the support from  University of Hyderabad through the IoE project grant IoE/RC1/RC1-20-012.


\begin{thebibliography}{}
\bibitem{Esteban:2020cvm}
I.~Esteban, M.~C.~Gonzalez-Garcia, M.~Maltoni, T.~Schwetz and A.~Zhou,
``The fate of hints: updated global analysis of three-flavor neutrino oscillations,''
JHEP \textbf{09}, 178 (2020)
[arXiv:2007.14792 [hep-ph]].

\bibitem{T2K:2021xwb}
K.~Abe \textit{et al.} [T2K],
``Improved constraints on neutrino mixing from the T2K experiment with $\mathbf{3.13\times10^{21}}$ protons on target,''
Phys. Rev. D \textbf{103}, no.11, 112008 (2021)
[arXiv:2101.03779 [hep-ex]].

\bibitem{NOvA:2021nfi}
M.~A.~Acero \textit{et al.} [NOvA],
``An Improved Measurement of Neutrino Oscillation Parameters by the NOvA Experiment,''
[arXiv:2108.08219 [hep-ex]].

\bibitem{DUNE:2020ypp}
B.~Abi \textit{et al.} [DUNE],
``Deep Underground Neutrino Experiment (DUNE), Far Detector Technical Design Report, Volume II: DUNE Physics,''
[arXiv:2002.03005 [hep-ex]].

\bibitem{Abazajian:2012ys}
K.~N.~Abazajian, M.~A.~Acero, S.~K.~Agarwalla, A.~A.~Aguilar-Arevalo, C.~H.~Albright, S.~Antusch, C.~A.~Arguelles, A.~B.~Balantekin, G.~Barenboim and V.~Barger, \textit{et al.}
``Light Sterile Neutrinos: A White Paper,''
[arXiv:1204.5379 [hep-ph]].

\bibitem{Berryman:2015nua}
J.~M.~Berryman, A.~de Gouv\^ea, K.~J.~Kelly and A.~Kobach,
``Sterile neutrino at the Deep Underground Neutrino Experiment,''
Phys. Rev. D \textbf{92}, no.7, 073012 (2015)
[arXiv:1507.03986 [hep-ph]].

\bibitem{Gandhi:2015xza}
R.~Gandhi, B.~Kayser, M.~Masud and S.~Prakash,
``The impact of sterile neutrinos on CP measurements at long baselines,''
JHEP \textbf{11}, 039 (2015)
[arXiv:1508.06275 [hep-ph]].

\bibitem{Agarwalla:2016xxa}
S.~K.~Agarwalla, S.~S.~Chatterjee and A.~Palazzo,
``Physics Reach of DUNE with a Light Sterile Neutrino,''
JHEP \textbf{09}, 016 (2016)
[arXiv:1603.03759 [hep-ph]].

\bibitem{Agarwalla:2016xlg}
S.~K.~Agarwalla, S.~S.~Chatterjee and A.~Palazzo,
``Octant of $\theta_{23}$ in danger with a light sterile neutrino,''
Phys. Rev. Lett. \textbf{118}, no.3, 031804 (2017)
[arXiv:1605.04299 [hep-ph]].

\bibitem{Coloma:2017ptb}
P.~Coloma, D.~V.~Forero and S.~J.~Parke,
``DUNE Sensitivities to the Mixing between Sterile and Tau Neutrinos,''
JHEP \textbf{07}, 079 (2018)
[arXiv:1707.05348 [hep-ph]].

\bibitem{Choubey:2017cba}
S.~Choubey, D.~Dutta and D.~Pramanik,
``Imprints of a light Sterile Neutrino at DUNE, T2HK and T2HKK,''
Phys. Rev. D \textbf{96}, no.5, 056026 (2017)
[arXiv:1704.07269 [hep-ph]].

\bibitem{Choubey:2017ppj}
S.~Choubey, D.~Dutta and D.~Pramanik,
``Measuring the Sterile Neutrino CP Phase at DUNE and T2HK,''
Eur. Phys. J. C \textbf{78}, no.4, 339 (2018)
[arXiv:1711.07464 [hep-ph]].

\bibitem{Choubey:2016fpi}
S.~Choubey and D.~Pramanik,
``Constraints on Sterile Neutrino Oscillations using DUNE Near Detector,''
Phys. Lett. B \textbf{764}, 135-141 (2017)
[arXiv:1604.04731 [hep-ph]].

\bibitem{Gandhi:2017vzo}
R.~Gandhi, B.~Kayser, S.~Prakash and S.~Roy,
JHEP \textbf{11}, 202 (2017)
doi:10.1007/JHEP11(2017)202
[arXiv:1708.01816 [hep-ph]].

\bibitem{DUNE:2020fgq}
B.~Abi \textit{et al.} [DUNE],
``Prospects for beyond the Standard Model physics searches at the Deep Underground Neutrino Experiment,''
Eur. Phys. J. C \textbf{81}, no.4, 322 (2021)
[arXiv:2008.12769 [hep-ex]].

\bibitem{DUNE:2015lol}
R.~Acciarri \textit{et al.} [DUNE],
``Long-Baseline Neutrino Facility (LBNF) and Deep Underground Neutrino Experiment (DUNE): Conceptual Design Report, Volume 2: The Physics Program for DUNE at LBNF,''
[arXiv:1512.06148 [physics.ins-det]].

\bibitem{Huber:2004ka}
P.~Huber, M.~Lindner and W.~Winter,
``Simulation of long-baseline neutrino oscillation experiments with GLoBES (General Long Baseline Experiment Simulator),''
Comput. Phys. Commun. \textbf{167}, 195 (2005)
[arXiv:hep-ph/0407333 [hep-ph]].

\bibitem{Huber:2007ji}
P.~Huber, J.~Kopp, M.~Lindner, M.~Rolinec and W.~Winter,
``New features in the simulation of neutrino oscillation experiments with GLoBES 3.0: General Long Baseline Experiment Simulator,''
Comput. Phys. Commun. \textbf{177}, 432-438 (2007)
[arXiv:hep-ph/0701187 [hep-ph]].

\bibitem{DUNE:2021cuw}
B.~Abi \textit{et al.} [DUNE],
``Experiment Simulation Configurations Approximating DUNE TDR,''
[arXiv:2103.04797 [hep-ex]].

\bibitem{DUNE:2021tad}
A.~Abed Abud \textit{et al.} [DUNE],
[arXiv:2103.13910 [physics.ins-det]].

\end{thebibliography}
\end{document}